\documentclass{article}    

\usepackage{graphicx}

\title{\textbf{Consciousness and Endogenous State Reduction: Two Experiments}}  
\author{Richard Mould\footnote{Department of Physics and Astronomy, State University of New York, Stony Brook,
\mbox{New York} 11794-3800; http://nuclear.physics.sunysb.edu/ \~{}mould}}  
\date{}    
   
\begin{document}             

\maketitle              

\begin{abstract}

      	There is a tradition in science that regards consciousness as merely epiphenomenal.  Accordingly, physical
systems can create and influence consciousness, but consciousness can have no influence on physical systems. 
Indeed, the current understanding of quantum mechanics provides no way for consciousness to alter the wave function
of a quantum mechanical state.  Furthermore, there is nothing in molecular biology that would suggest that the human
body is anything more that an automaton that operates on the basis of purely physical and chemical interactive
forces.  However, I believe that the epiphenomenal view is fundamentally flawed, and I suggest the following
experiments as a way of demonstrating the existence of an influence of consciousness on material systems.  The first
uses Positron Emission Tomography (PET) with a human subject, and the second used autoradiography with rats.

Detailed arguments for my position can be found in three papers that have been published in recent years.  A brief
summery of the arguments is initially given below, where it is claimed that `pain' consciousness might be correlated
in a certain way with the relative binding of opiates to receptors in a subject's brain.  

\end{abstract}

\section*{1. Introduction}

 In recent papers\cite{RM1}\cite{RM2}\cite{RM3}, I accept the evolutionary argument of William James to the effect
that psychological states must have evolved along with the biological states\cite{WJ}. For these two very different
kinds of things to have evolved in parallel with one another (i.e., for one to have anything to do with the other),
James says that there must have been an interaction between the two.  Otherwise, wrong psychological constructions
would not have been selected against, and therefore, would have improperly survived the evolutionary struggle.  If
that had been the case, then any resemblance between the subjective imagery of our species and the world about us
would be completely fortuitous.  This argument is given more fully in ref.\ 2.  It says that the psycho-physical
parallelism of von Neumann must have come about through a natural process in which psychological and physical states
were mutually engaged.  Otherwise, one would have to believe either in Libnitz's claim of a (miraculous)
\emph{pre-established harmony} between these two things, or in Bishop Berkeley's denial that there exists a
psycho-physical parallelism in the first place.  

According to von NeumannÕs interpretation of quantum mechanical measurement, the collapse of the wave function
requires the presence of a subjective (i.e., conscious) observer\cite{vN}. I make use of this idea together with the
notion of an \emph{inside observer} (initially defined in ref.\ 1) to show how conscious states arising within a
physical system might conceivably influence the probability amplitude of quantum mechanical choices made by the
system.  An inside observer is defined to be a \emph{state of conscious awareness} that emerges on one component of
an endogenous quantum mechanical superposition of (physiological) states. These states can be macroscopic in the
formalism of quantum mechanics; although, in deference to environmental entanglement and decoherence, one might call
them ``mixtures" instead of ``superpositions".  I do not use this language because I am not concerned here with
coherence or interference between macrostate components.\footnote{Environmentally entangled macrostates really are
superpositions.  Joos and Zeh say, ``... the interference terms still exist, but they are
not \emph{there}."\cite{JZ}  This paradoxical statement means that the system's phases exist between global states
that include non-local correlations connecting a macrostate with its environment.  These phases are not accessible
to a local observer; and in consequence, the superposition appears locally to be a mixture.\cite{Gea}  But however
one regards such a macrostate ensemble, as a global superposition or as a local mixture, each component has a
probability amplitude in a fully quantum mechanical system.  Therefore, according to von Neumann, a conscious
observation is necessary for one of these states to become a concrete reality.}  

When different inside observers finally do emerge on different components of a physiological superposition, I
propose that there will be a shift in the relative probability amplitudes of the components that favors certain
conscious states over others.  The principles that govern this leaning toward especially favored states are
described below.  I assume that the above shift preserves normalization.\footnote{I do not say that this shift among
probability amplitudes is \emph{caused} by the appearance of consciousness, inasmuch as the underlying influence is
revealed only as an empirical relationship.  The existence of various conscious states in the endogenous
superposition, \emph{plus} my hypothetical change in relative probability amplitudes, \emph{plus} an accompanying
collapse of the wave function may all result from a `common' unknown cause\cite{RM4}.}

This possibility provides us with an opening in quantum mechanics that may admit the evolutionary influence
envisioned by James.  What is needed is a model of some primitive species at the time of its first use of
consciousness, together with an identification of the kind of conscious experience that can influence the creature's
evolution by using the above `inside observer' mechanism.

\section*{2. My Hypothesis}

I believe that the first conscious experiences that appeared in any evolving species must have been very
straightforward, like simple \emph{pleasure or pain}.  Elemental perceptions involving sight, sound, or touch serve
no behavioral purpose in themselves, for they have no intrinsic motivational weight or direction.  These perceptions
must be `interpreted' in order to have significance, and that is too much to expect of the first glimmer of
consciousness.  In addition, emotions such as fear, anger, and love have no intrinsic meaning apart from an existing
subjective construction of the world toward which they are directed.  This means that an awareness of either one of
these emotions requires a greater sophistication than is required for simple experiences like pleasure or pain.  The
latter have a direct `unsophisticated' motivational power that stands apart from any concept of the external
world.\footnote{This excludes `emotional' pain.  It refers only to `physical' pain (e.g., a flesh wound or a broken
bone)}   I therefore develop an evolutionary model that uses `pain' as a creatureÕs first conscious experience,
where pain consciousness is said to have an influence on the quantum mechanical choices that are made within the
creature.  This is done in detail in ref.\ 2, and refined in ref.\ 3.  

My hypothesis requires that if an endogenous quantum mechanical superposition develops within a conscious creature
in which a more painful component competes with a less painful one, than, other things being equal, \emph{the less
painful component will have an enhanced probability of surviving a collapse of the state function}.  This is the
interactive mechanism whereby conscious states are claimed to influence matter.  The model assumes that the more
painful experience is associated with a life-threatening behavior in a way that is best explained in ref.\ 3, pp.
1953-4.  

An endogenous quantum mechanical superposition develops within the experimental subjects because the
\emph{ligands} that attach to receptors in the brain have quantum mechanical wave functions that spread rapidly in
space due to the Heisenberg uncertainty principle.\footnote{The term ligand refers to any molecule, endogenous or
exogenous, that attaches to a receptor.}  The spreading takes place as ligands are swept along in blood and
cerebrospinal fluid on their way to the receptor.  This means that there is an intrinsic probability governing the
number of ligands that become attached to the opiate receptors in a pain responsive region of the brain, thereby
posing a quantum mechanical choice between a more painful experience and a less painful experience.  `\emph{More
pain' and `less pain' are eigenvalues of the endogenous state reduction}.\footnote{Another way of saying this is
that there is a `more pain' observer, and a `less pain' observer, who are competing ``inside observers" associated
with different components of the endogenous quantum mechanical superposition.  The reduction is one that makes a
definite choice between these contenders}   In this situation, my hypothesis requires an increase in the probability
of the less painful eigenstate.  One would therefore expect to find a surviving endogenous eigenstate to contain more
opiate-like molecules in a part of the brain that mediates pain, than would be expected on the basis of biochemical
considerations alone.  Most measurements of ligand bonding in humans appear to have been made using
subpharmacological doses, so it is clear that the above hypothesis needs to be tested in vivo using doses that are
large enough to be felt.  That is the purpose of these experiments. 

Ligands are called \emph{agonists} if they produce cellular effects within the receptor to which they are attached. 
Morphine, fentanyl, and carfentanil are examples of opiate receptor agonists, inasmuch as they cause cellular
disturbances that we recognize as analgesia and/or euphoria.  Other molecules produce no pharmacological effects
when they are attached to receptors.  These are called \emph{antagonists}.  Naloxone and diprenorphine are examples
of opiate receptor antagonists.  When a mixture of an agonist and an antagonist is injected into a subject, the two
molecules will compete with on another for attachment to the available receptors.  Such a mixture (in sufficient
dose) will produce pharmacological effects, owing to the attached agonist molecules.   

The competition between the two substances is expressed quantum mechanically by their appearing in different ratios
on different components of the endogenous superposition. Competing components of the superposition will therefore
support competing inside observers who experience different degrees of pain.   It is my claim that the probability
amplitudes of these components will be skewed in favor of the observer experiencing less pain.\footnote{My
hypothesis chooses less pain to be favored over more pain for reasons that are best understood in terms of the
evolutionary model developed in refs. 2 and 3.}

No attempt is made here to determine which agonists and antagonists are best suited for the experiments. 
Carfentanil and diprenorphine may be acceptable for the experiment with rats; but the toxicity of carfentanil makes
is unsuitable for use with humans in pharmacological doses.  Perhaps fentanyl and naloxone would be the best
combination for humans.  However, there may be no ideal choice of ligands at this time.  Background radiation may
swamp our anticipated results because of ligand binding that is not sufficiently specific to the targeted receptors,
and/or because of the existence of too many free unbound ligands.  Research is ongoing to find ligands with greater
affinity and specificity.  Therefore, although the following experiments may not now give unambiguous results, they
can be thought of as idealized experiments to be performed when the technology is sufficiently improved.

\section*{3. Preliminary to the Experiments}

Prior to one of the experiments, the subject who is exposed to painful trauma should be given a prescribed mixture
of an agonist and an antagonist to determine the size of the dose that brings the subject to the threshold of
analgesia.  It is assumed that a \emph{threshold dose} is small enough that opiate receptors remain unsaturated
throughout the brain, and in fact, that the number of bound receptors in each region remains linear with the number
of ligands that are available to the receptors.

\section*{4. The First Experiment}

	Four PET scans are proposed that will allow a comparison to be made between the binding of an agonist and an
antagonist in the brain of a human suffering from rheumatoid arthritis (or other chronic pain), or one who is
subjected to cutaneous applications of a pain producing heat (or other inflicted pain).  Each scan begins with an
intravenous injection of agonist and antagonist in ratio $R$.  This ratio is chosen to insure that the number of
agonist molecules that become attached to the receptors is roughly equal to the number of antagonist molecules that
become attached to the receptors.  Only one of these substances is labeled radioactively during a single scan.   A
scan might reasonably begin 30 minutes after injection and last for 45 minutes. 

	In the first scan, the subject is given a threshold dose of hot agonist and cold antagonist.  After the scan, the
receptor count/pixel given by $C_A$ is recorded in each ROI (neurological region of interest).  In the second scan,
the subject is given a threshold injection of cold agonist, and hot antagonist, which has a net weight equal to that
of the first injection.  After that scan, the receptor count/pixel, given in this case by $C_{AA}$, is recorded in
each ROI.  The total receptor count/pixel is then $C = C_A + C_{AA}$, and the ratio is 

\begin{equation}
r = C_A/C_{AA}
\end{equation}
in each ROI.  If the test subject is required to endure cutaneous heat, then this will be applied from the time of
injection to the end of the scan.  

	The first row in fig.\ 1 represents the population of molecules that occupy opiate receptors in some ROI in the
first scan.  The agonist is shown to be radioactive in the first row (as indicated by the wings), where the number
of endogenous ligands (e.g., endorphins or other peptides) that are competing for site receptors is indefinite.  

	The second row of fig.\ 1 represents the population of molecules that occupy the opiate receptors in the second
scan, where the antagonist molecules are now radioactive.  Combining the first and second scans allows one to
measure $C_A$ and $C_{AA}$ in each ROI, and so to find $C$ and $r$.  

\begin{figure}[t]
\centering
\includegraphics[scale=0.8]{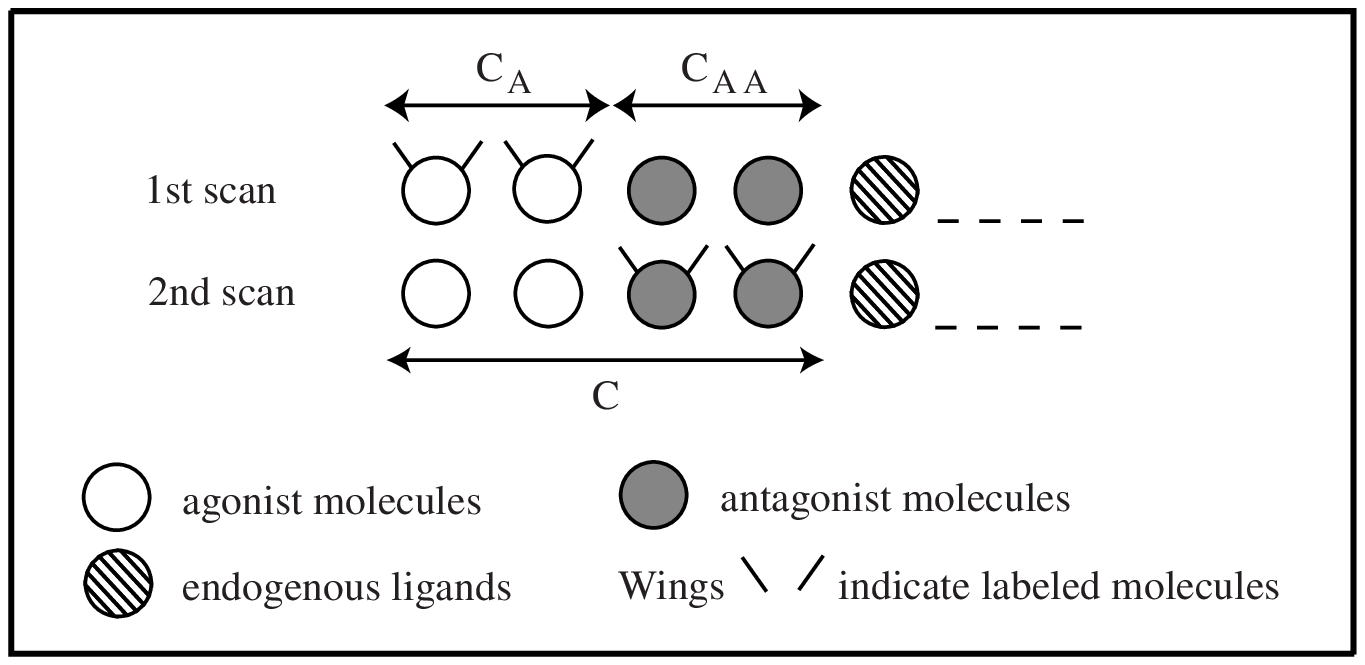}
\center{Figure 1}
\end{figure}

	The third and fourth scans are identical with the first and second, except that the doses in this case are
subpharmacological.  This guarantees that consciousness will have nothing to do with the result.  The third and
fourth scans therefore provide values of $r$ for each ROI that are determined by all known biochemical
influences as well as any purely methodological influences.  It is my assumption that 
\begin{eqnarray}
r&\mbox{(pain responsive ROIs in scans one and two)}\\
>r&\mbox{(pain responsive ROIs in scans three and
four)}\nonumber
\end{eqnarray}
and that $r$ will be the same in regions that are not pain responsive.  The only thing that distinguishes these two
values of $r$ in eq.\ 2 is the size of the dose, and biochemically speaking, that should not have anything to with
the result. That's because the only biochemical influence on the observed value of $r$ is the competition between the
agonist and the antagonist, \emph{and dose should not affect this balance for small doses at steady state
equilibrium.}\footnote{Small dose means: on the linear portion of the binding $vs$ concentration curve, far from
saturation\cite{MT}.  For ligands of normal potency, this should present no difficulty for threshold doses.}  The
inequality in eq.\ 2 is therefore a test of my hypothesis concerning the influence of pain consciousness.  It
suggests that there are non-biophysical influences operating in the pain responsive regions of the brain that result
in more agonist molecules being bound to these regions than would otherwise be expected.  Presumably, this is because
the conscious observer (i.e., the PET subject) becomes associated with a collapse of an endogenous wave function that
gives preferential weight to less painful eigenstates.  In order to isolate the observer in this experiment,
monitors carrying raw data from the scanner should be covered during the time of the scan.

\section*{5. Distribution in r}

The ratio $r = C_A/C_{AA}$ is a variable of the total endogenous quantum mechanical state.  $C = C_A + C_{AA}$ is
another variable, but it is of no interest here.  

The pulse appearing on the left in fig.\ 2 represents the distribution of eigenstates of $r$ in a region of the brain
that is not responsive to pain.  These eigenstate amplitudes will be the same as those predicted by quantum
physiology.  

In regions that are responsive to the pain, the eigenstates on the right-hand slope of the left-hand pulse will have
proportionally more agonist molecules than those that are on the left-hand slope, inasmuch as they represent states
having a greater ratio $r$.  So eigenstates on the left-hand slope are more painful  than  those on the right-hand
slope.  Therefore, according to my hypotheses, the eigenstates on the right-hand slope will grow in amplitude
relative to those on the left.  When this process comes to equilibrium, the entire pulse will have displaced to the
right as shown in fig.\ 2 where its components will be richer in the analgesic agonist.  The measured value of $r$
will therefore be greater in these regions.  

The quantum mechanical state function for the subject's entire body is a function of many variables including $r_A,
r_B, r_C, r_D,$ . . . etc., where these represent the ratio $r$ in regions $A, B, C, D,$ . . . etc.    The total
physical state can therefore be written in the form $\Psi(r_A, r_B, r_C, r_D,$ . . . etc.).  Presumably the conscious
state of the subject is determined by $\Psi$ in its entirety.  The distribution of $r$ in regions that are not pain
responsive will be determined by the quantum mechanics alone.  However, displacements in the functional dependence
of $r$ in all of the regions that are pain responsive will decrease the pain consciousness of the organism of the
whole.    

\begin{figure}[t]
\centering
\includegraphics[scale=0.7]{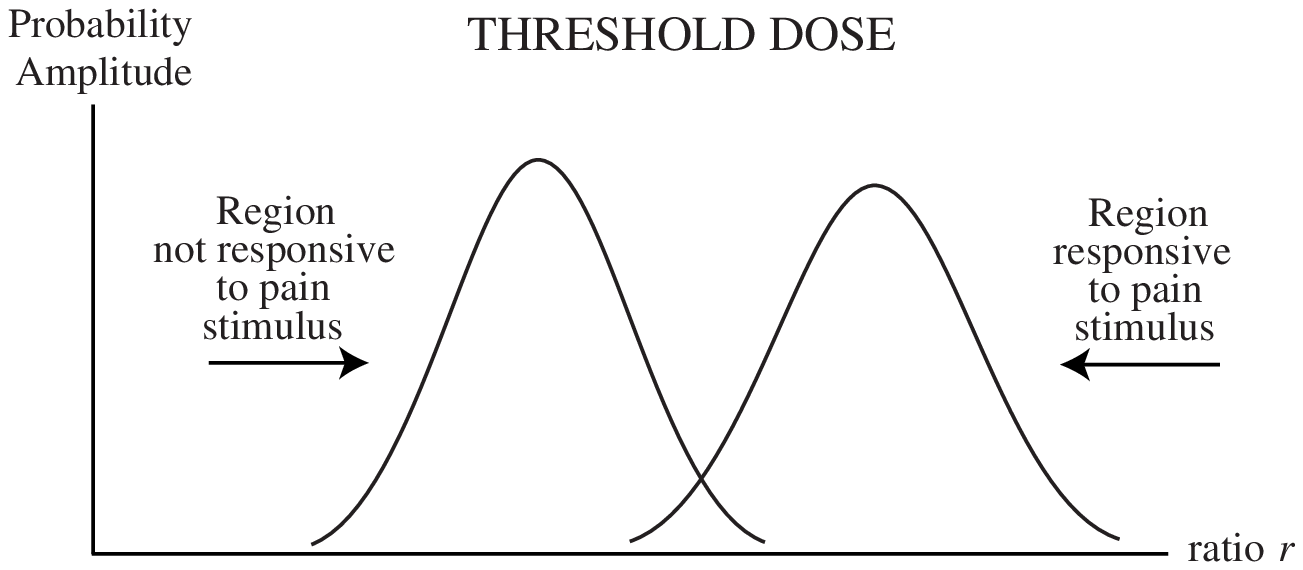}
\center{Figure 2}
\end{figure}

\section*{6. Discussion}

	In pain responsive ROIs, endogenous opioid agonists will generally be secreted as part of an attempt by the body to
alleviate the pain.  As a result, the total number C of exogenous ligands will be decreased in these regions as has
been shown in other studies\cite{AJ}.  However, this decrease will not matter to the experiment because it is the
ratio $r$ of the two injected ligands that is important.  That ratio is governed only by the competition between the
two, and is independent of other secretions.  This is one reason why the experiment is based on the unitless ratio
$r$. A concern is that at pharmacological doses, this effect will result in a more homogeneous radioactive response
over all regions of the brain, which would tend to mask differences in $r$.  

	The experiment will be valid, even if the agonist is specific to, say $\mu$-receptors, and the antagonist is
non-specific, as is the case of carfentanil and diprenorphine.  The ratio $r$ should certainly be affected by
variations of specificity, but it would remain the same between the first two scans and the second two scans in this
experiment.  Whatever the value of $r$ for subpharmacological doses, it should not change (biochemically speaking) as
the dose is increased to threshold, even if the ligands engage different populations of receptors.  A concern is
that differences in $r$ will be masked by excessive non-specificity. 

	It will also not matter if the size of the dose is slightly different for different subjects.  A different dose
will change the total count C in each ROI, but the ratio $r$ in each region should not thereby be affected.  What is
important is that the dose be one that puts the injected agonist and antagonist molecules into one-on-one
competition with one another at the analgesic threshold of the subject.  

	Since it is impossible at this point to estimate the magnitudes of the hypothetical displacement of the pulse in
fig.\ 2, it is impossible to estimate the number of data points that would be necessary to get good statistics. 
Resolution is part of what must be decided by the experiment.

\section*{7. The Second Experiment}

	There is another approach to this problem that involves in vivo autoradiography with rats experiencing pain. 
Instead of four PET scans, there are four injections (in four different rats) of a mixture of agonist and antagonist
that follow the same protocol as before.  Each of the four rats is sacrificed.  The brain of each is then sliced and
exposed to film to reveal the concentration of labeled ligands in different parts of the brain.  

	As in the PET case, the first two doses will allow a determination of $r = C_A/C_{AA}$ in each ROI at threshold
levels.  The second two doses will allow $r$ to be determined in each ROI at subpharmacological doses, which insures
that consciousness will not be a factor.  My claim is that eq.\ 2 should also apply in this case, giving evidence of
the influence of consciousness on the outcome.   

The second (autoradiographic) experiment might be cheaper and easier than the first (PET) experiment. 
But there is one great disadvantage.  A negative result might mean that rats are automatons that have no conscious
life.  It is only in a truly in vivo experiment (i.e., one in which the subject is alive and known to be fully
conscious when the data is taken) that the hypothesis can be fully tested.  This requires a PET based experiment
that uses human subjects.  On the other hand, a positive result for the second (autoradiographic) experiment would
be useful because it would suggest that rats are conscious, and that my hypothesis is correct.

\section*{Acknowledgements}

		I would like to thank Ron Blasberg, Richard Carson, Joanna Fowler, James Frost, Anthony Jones, and Milt Titeler
for helping me to understand some of the capabilities and limitations of PET technology and autoradiography.  I
appreciate their willingness to share their extensive experience with me in the field of opiate ligands and their
receptors.

\end{document}